\documentclass[superscriptaddress,amsmath,amssymb]{revtex4}
\usepackage{graphicx}
\begin{document}
\title{Two neutron correlations in exotic  nuclei}

\author{H. Sagawa}
 \affiliation{Center for Mathematics and Physics, University of Aizu,
Aizu-Wakamatsu, 965-8580 Fukushima, Japan},
%%  \email={sagawa@u-aizu.ac.jp},
%%}

\author{K. Hagino}{
\affiliation{Department of Physics, Tohoku University, Sendai, 980-8578,  Japan},
%% \email={jmargue@ipno.in2p3.fr},
%%  homepage={http://www.dcarlisle.demon.co.uk},
%%}

\title{Two neutron correlations in exotic  nuclei}
\date{}
%%\classification{21.30.Fe,21.60.Jz,21.65.-f,26.50.+x}

\begin{abstract}
We study the correlations between two neutrons    
% KH*****
%correlations 
% KH*****
in borromian nuclei  $^{11}$Li and $^6$He
 by using a three-body model with a density-dependent 
  contact two-body interaction.
It is shown that the two neutrons show a compact bound feature at the nuclear 
 surface due to the mixing of single particle states of different parity. 
%%%%%%%% KH
% surface due to the mixing of different parity single particle states.
%%%%%%%% KH
We study the Coulomb breakup cross sections of   $^{11}$Li and $^6$He 
using the same three-body model.   
We show that the concentration of the B(E1) strength 
near the threshold can be well reproduced  with this model
as a typical nature of the halo nuclei. 
 The energy distributions of two emitted neutrons from
dipole excitations are  also studied  using the correlated 
wave functions of dipole excitations.
\end{abstract}

\vspace{-1cm}
\maketitle

%%\tableofcontents

%%\bigskip

%%\section{Introduction}
It has been well recognized by now that the borromian nuclei such as 
$^{11}$Li and  $^6$He show a strong di-neutron correlations in 
the ground states and also in the excited states.
%%\cite{BW52,BBH91,STG92,NLV05,TB05}. 
Recently, Nakamura {\it et al.} have remeasured the low-lying 
dipole excitations in $^{11}$Li nucleus and have confirmed 
   the strong concentration of the dipole 
strength near the threshold in the  2-neutron (2n) halo nucleus
\cite{N06}. The low-lying dipole strength for another
2n halo nucleus, $^6$He, has also been measured by 
Aumann {\it et al.} \cite{A99}. The two neutron correlations are further 
 measured very recently 
  in the Coulomb breakup process of dipole excitaitons in $^{11}$Li
 \cite{Nakamura09}. 

The aim of this paper is to study the correlations between di-neutrons and 
also neutron-core correlations in borromian nuclei $^6$He and $^{11}$Li
 by using a three-body model \cite{BF1,BF2,HS05}.
In Ref. \cite{HSCS07}, 
 the behavior of the two valence neutrons 
in $^{11}$Li  is studied 
at various positions from the center to the surface of the nucleus. 
It was  found that  the two-neutron wave function 
oscillates near the center whereas it becomes similar to 
that for a compact bound state around the nuclear surface,
and  the mean distance between the valence neutrons has a
well pronounced minimum around the nuclear surface.
%% as is shown in Fig. 
%%\ref{fig:correlation-length}. 
We have pointed out that 
these are 
qualitatively the same behaviors as found in neutron matter \cite{M06}. 
To elucidate these points, we show in Fig. \ref{fig:correlation-length} the 
mean distance of the valence neutrons 
in $^{11}$Li
as a function of 
the nuclear radius  $R$  (the distance between the core and the center of two neutrons)  obtained with and without the 
neutron-neutron ($nn$) interaction. 
For the uncorrelated calculations, we consider both the
[(1p$_{1/2})^2$] and [(2s$_{1/2})^2$] configurations. 
One can see that, in the non-interacting case, 
the neutron  pair almost monotonously 
expands, as it gets further away from the center of the nucleus. 
On the other hand, in the interacting case it first becomes  smaller  going from 
inside to the surface before expanding again into the free space 
configuration. These results confirm  the strong and predominant influence
 of the pairing force 
in the nuclear surface of$ ^{11}$Li. We also show the local 
 neutron-neutron correlation energy as a function
of the radius $R$ in the lower 
panel of Fig. \ref{fig:correlation-length}.
It is clearly seen that the energy gain is the maximum at the surface where 
 the correlaiton length is the minimum. 
 The two  panels in Fig. \ref{fig:correlation-length} 
 confirm that the kink of the correlation length is induced by the 
 strong pairing correlations at the surface.

%%\vspace{-1cm}
\begin{figure}
\includegraphics[scale=0.35,clip]{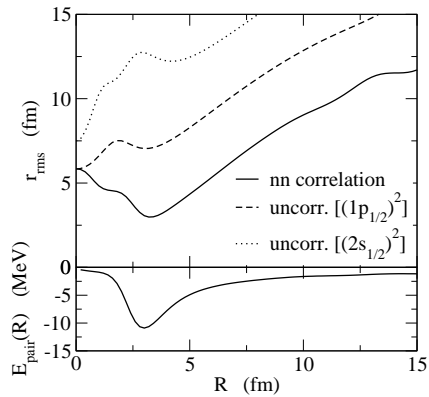}
\caption{
%%(Color online)
(Upper panel)The root mean square distance $r_{\rm rms}$ between the valence neutrons in
$^{11}$Li as a function
of the distance $R$ between the core and the center of two neutrons. 
The solid line is obtained by taking into account the neutron-neutron
 correlations,
while the dashed and the dotted lines are obtained 
by switching off the neutron-neutron interaction and assuming the 
[(1p$_{1/2})^2$] and [(2s$_{1/2})^2$] configurations,
respectively. 
(Lower panel) The neutron-neutron correlation energy as a function
of the distance  $R$.
\label{fig:correlation-length}}
\end{figure}
%%\vspace{-1cm}

\begin{figure}
\includegraphics[scale=1.1,clip]{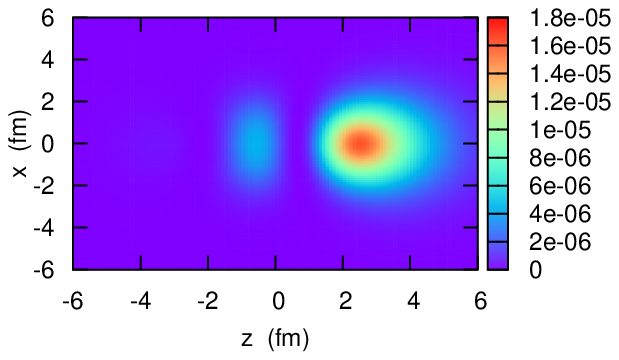}
\includegraphics[scale=1.1,clip]{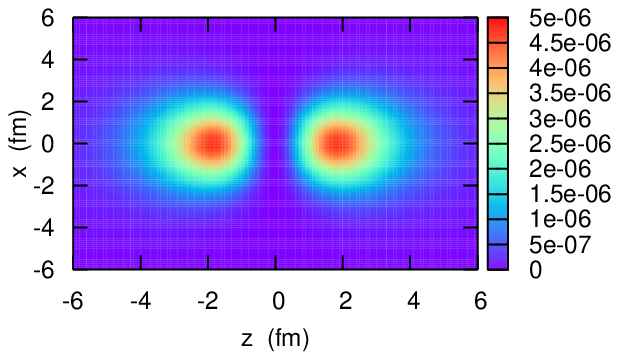}
\caption{
(Color online)
Two dimensitional (2D) plots for the two particle density of  the
correlated pair (left panel) and uncorrelated [(1p$_{1/2})^2$]  configuration
 (right  panel) in $^{11}$Li.  
  It represents the probability distributions for the 
spin-up neutron placing the spin-down neutron at $(z,x)$=(3.4,0)fm.  
The core  nucleus is located at the origin $(z,x)$=(0,0)fm.
\label{fig:2Dplot}}
\end{figure}
%%\vspace{-1cm}

 Two particle densities of the  correlated pair  and the uncorrelated
  [(1p$_{1/2})^2$] configuration are 
 shown in Fig.~\ref{fig:2Dplot}.
   The reference particle is located at $(z,x)=(3.4,0)$fm.  
As can be seen in the  right panel, the distribution has a symmetric two peaks in 
  $(z,x)$ plane with respect to the center of the core nucleus at
 $(z,x)=(0,0)$fm.  This is due to the absence of mixing of opposite parity 
wave functions into the  [(1p$_{1/2})^2$] configuration.  
 On the contrary, the peak appears only around the position of the reference 
 particle when the two neutron  correlations are taken into account in the wave
 functions.  To compare two panels in Fig. \ref{fig:2Dplot}, we can see
a  clear manifestation of the strong two neutron 
correlations in the wave function of the borromian nucleus $^{11}$Li.

\begin{figure}
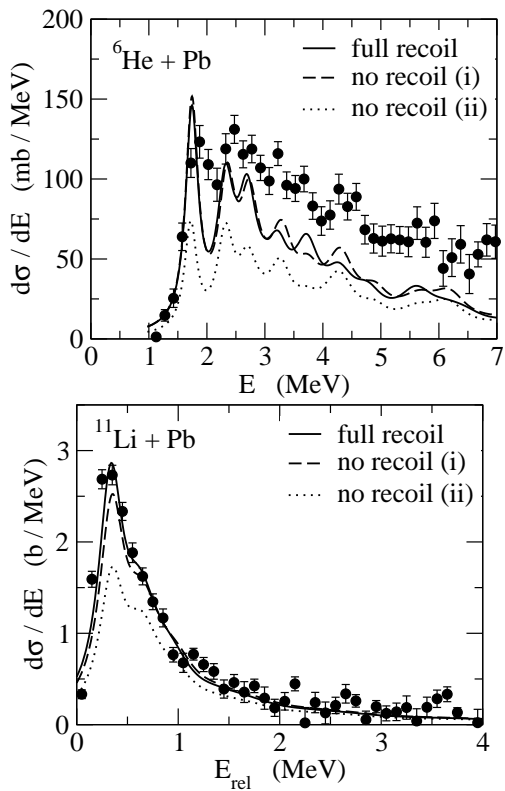

\includegraphics[scale=0.5,clip]{fig3-1dub}\\
\includegraphics[scale=0.5,clip]{fig3-2dub}\\
\caption{
Coulomb breakup cross sections for $^6$He+Pb at 240 MeV /nucleon and
for the $^{11}$Li +Pb at 70 MeV/nucleon. 
The solid line is the result of the full  calculations,
while the dashed line is obtained by neglecting the off-diagonal
component of the recoil kinetic energy in the excited states. 
The dotted line is obtained by neglecting the off-diagonal recoil term
both in the ground and the excited states. 
These results are smeared with an energy dependent width of 
$\Gamma = 0.15 \cdot \sqrt{E_{\rm rel}}$ MeV for $^6$He and 
$\Gamma = 0.25 \cdot \sqrt{E_{\rm rel}}$ MeV for  $^{11}$Li. 
The experimental data are taken from Refs. \cite{A99} and \cite{N06}
for  $^6$He and  $^{11}$Li, respectively.
\label{fig1}
}
\end{figure}

Figs. \ref{fig1}  compare 
the Coulomb breakup cross sections calculated 
 by taking into account  
the recoil term exactly (the solid curves) with those calculated approximately (the dashed 
and dotted curves). 
For the dashed curves, the off-diagonal component of the recoil kinetic
energy is neglected in the excited $J^\pi=1^-$ states, while it is fully taken 
into account in the ground state. 
It is interesting to notice that these calculations lead to similar 
results to the one in which the recoil term is treated exactly (the
solid curves).   
The dotted curves, on the other hand, are obtained by neglecting the 
off-diagonal part of the recoil term both for the ground and the 
$J^\pi=1^-$ states. 
%%For this calculation, we slightly readjust the
%%parameters of the pairing interaction so that the ground state energy 
%%remains the same. 
By neglecting the recoil term in the ground state, 
the value for $\langle r^2_{c-2n}\rangle$ decreases, 
from 13.2 fm$^2$ to 9.46 fm$^2$ for $^6$He 
and 
from 26.3 fm$^2$ to 20.58 fm$^2$ for $^{11}$Li. Consequently, 
the B(E1) distribution as well as the breakup cross sections are
largely underestimated. 
%%The fraction of the main components in the ground
%%state wave function are also altered by neglecting the recoil term: 
%%for $^6$He, the fraction of the (p$_{3/2})^2$ component changes from 
%%83.0 \% to 90.8 \%, 
%%and for $^{11}$Li, the fraction of the (s$_{1/2})^2$ component changes from 
%%22.6\% to 17.1\%, 
%%and the fraction of the (p$_{1/2})^2$ component from 
%%59.1 \% to 65.7 \%. 
 These results clearly indicate that the 
recoil term is important for the ground state, 
 while it has a rather small effect on the excited states.
%although it may be
%neglected in the excited states. 

\begin{figure}[htb]
\hspace{-1cm}
\includegraphics[clip,scale=1.5]{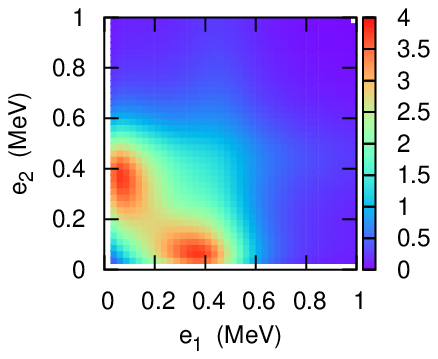}
\hspace{-1.5cm}
\includegraphics[clip,scale=1.5]{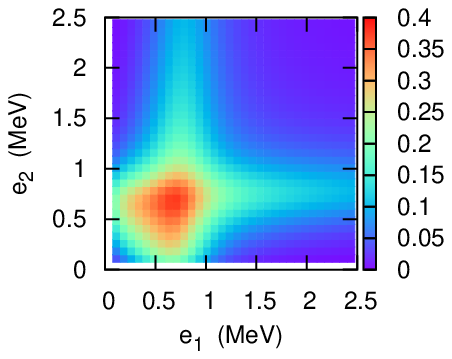}
\caption{(Color online)
The dipole strength distributions, $d^2B(E1)/de_1de_2$, of 
$^{11}$Li (left panel) and   $^6$He (right panel) as a function of
the energies of the two emitted neutrons relative to the core nucleus. 
They are plotted in units of $e^2$fm$^2$/MeV$^2$. 
Figures show  the correlated response, which 
fully takes into account the ground state 
 and final state interactions between the 
two neutrons.
\label{fig3}
}
\end{figure}

Figures \ref{fig3} show the dipole strength distribution, $d^2B(E1)/de_1de_2$, 
as a function of the energies of the two emitted neutrons for the 
$^{11}$Li and 
$^6$He nuclei, respectively \cite{Hagino09}.
One immediately notices that the strength distribution is considerably 
different between $^{11}$Li and $^6$He. For $^{11}$Li, a large concentration 
of the strength appears at about $e_1$=0.375 MeV and $e_2$=0.075 MeV 
(and at $e_1$=0.075 MeV and $e_2$=0.375 MeV), with 
a small ridge at an energy of about 0.5 MeV. 
On the other hand, for $^6$He, the strength is largely concentrated 
  at one peak around $e_1=e_2=0.7$ MeV and 
only a large ridge at about 0.7 MeV appears. 
This difference between    $^{11}$Li and  $^6$He is due to the existence of 
virtual s-state in the residual $^{10}$Li, but not in $^5$He.
   Thus, if the interaction between the neutron and the core nucleus 
is  switched off,  the distribution become  
similar between the two nuclei because the virtual s-state is diappeared in 
$^{10}$Li.  
%%On the other hand, even if the final state interaction between the 
%%two emitted neutrons is discarded, 
%%  the  features of the energy distributions remain the same 
%%as the full calculations.  

In summary, we have studied the di-neutron correlations and the neutron-core 
correlations in the borromian nuclei $^6$He and $^{11}$Li by using 
  the three-body model with a density dependent 
contact interaction.  It is shown that the two neutron wave functions show a
 strong di-nuetron correlation at the nuclear surface 
 due to the mixing of different 
parity single particle  states. The same model is used 
 to analyze the dipole strength 
   distributions as well as the
Coulomb breakup cross sections of the $^6$He and $^{11}$Li nuclei. 
We have shown that the strong concentration of the B(E1) strength near 
the continuum threshold can be well reproduced as a nature of the halo nuclei 
 with the present model.  
%%We have also 
%%examined the recoil effect 
%%of the core nucleus on the Coulomb breakup cross sections.  
It is shown that
the recoil effect plays an important role
in the ground state while it may
be neglected in the excited states.  
 We have  carried out  the calculations of the energy and angular 
distributions of the two emitted neutron from E1 excitations in 
  $^{11}$Li and $^6$He nuclei. 
We have shown that these distributions are strongly 
affected by the existence of the virtual s-state in the residual $^{10}$Li 
nucleus.  Thus, 
the properties of 
the neutron-core potential is crucial to describe the energy distributions of 
two emitted neutrons, rather than the two neutron correlaions in
 the excited states.  
%%For the $^{11}$Li nucleus, the presence of $s$-wave virtual state 
%%helps to reveal a clear manifestation of the strong 
%%dineutron correlation through the energy and 
%%the angular distributions.
%%\medskip

This work was supported by the Japanese
Ministry of Education, Culture, Sports, Science and Technology
by Grant-in-Aid for Scientific Research under
the program numbers (C) 19740115 and 20540277.

\end{document}